\documentclass[12pt]{article}
\makeatletter
\@addtoreset{equation}{section}
\makeatother

\usepackage[pdftex]{graphicx}
\usepackage{subfigure}
\usepackage{array}
\usepackage{amsmath}

\usepackage{rotating}
\usepackage{longtable}
\usepackage[T1]{fontenc}
\usepackage{mathtools}
\usepackage{graphicx}
\usepackage{floatrow}
\usepackage{subfig}

\usepackage{float}
\usepackage{array}
\usepackage{geometry}
\usepackage{color}
\usepackage{lipsum}

\usepackage{diagbox} 
\usepackage{pdfpages}
\usepackage{multicol}
\usepackage[utf8]{inputenc}
\usepackage[english]{babel}
\usepackage{mathtools}
\usepackage{bbm}
\usepackage{amssymb}
\usepackage{amsthm}
\usepackage{hyperref}
\usepackage{bm}
\usepackage{ragged2e}
\usepackage{fancybox}
\usepackage{multirow}
\usepackage{authblk}
\usepackage{algorithm}
\usepackage{algpseudocode}
\usepackage{setspace}
\usepackage{caption}
\usepackage{fancybox}

\usepackage{adjustbox}
\usepackage[utf8]{inputenc}
\usepackage{tikz}

\allowdisplaybreaks

\usetikzlibrary{shapes.geometric,arrows}
\tikzstyle{startstop} = [rectangle, rounded corners, minimum width=3cm, minimum height=1cm,text centered, draw=black]
\tikzstyle{io} = [trapezium, trapezium left angle=70, trapezium right angle=110, minimum width=3cm, minimum height=1cm, text centered, draw=black]
\tikzstyle{process} = [rectangle, minimum width=3cm, minimum height=1cm, text centered, draw=black]
\tikzstyle{decision} = [diamond, minimum width=3cm, minimum height=1cm, text centered, draw=black]
\tikzstyle{arrow} = [thin,->,>=stealth]

\newlength\szg
\newcommand\quan[1]{%
\settoheight\szg{#1}%
\tikz[baseline]{\pgfmathparse{
    ifthenelse(#1 < 10, 1, ifthenelse(#1 < 100, 0.75, 0.5))
}
\let\hfs\pgfmathresult
\node at (0,\szg/2) {\makebox[0em][c]{\scalebox{\hfs}[1]{#1}}};
\draw (0,\szg/2) circle (\szg/2+0.35ex);
}}

\makeatletter
\newcommand*{\rom}[1]{\expandafter\@slowromancap\romannumeral #1@}
\makeatletter
\def\BState{\State\hskip-\ALG@thistlm}
\makeatother
\usepackage{xcolor}
\renewcommand{\raggedright}{\leftskip=0pt \rightskip=0pt plus 0cm}
\title{Shape-Preserving Prediction \\ for Stationary Functional Time Series}
\author[]{Shuhao Jiao\thanks{shjiaoqd@gmail.com} }
\author[]{Hernando Ombao\thanks{hernando.ombao@kaust.edu.sa}}

\affil[]{Statistics Program,  KAUST, Saudi Arabia}
\date{}
\allowdisplaybreaks
\begin{document}
	\maketitle			
	\setlength\parindent{0pt}
	\setlength{\parskip}{1em}
	\theoremstyle{definition}
	\newtheorem{thm}{Theorem}
	\newtheorem{lemma}{Lemma}
	\newtheorem{assumption}{Assumption}
	\newtheorem{prop}{Proposition}
	\newtheorem{definition}{Definition}
	\newtheorem{remark}{Remark}

\begin{abstract}
This article presents a novel method for prediction of stationary functional time series, in particular for trajectories that share a similar pattern but with variable phases. The limitation of most of the existing prediction methodologies for functional time series is that they only consider vertical variation (amplitude, scale, or vertical shift). To overcome this limitation, we develop a shape-preserving (SP) prediction method that incorporates both vertical and horizontal variation. One major advantage of our proposed method is the ability to preserve the shape of functions. Moreover, our proposed SP method does not involve unnatural transformations and can be easily implemented using existing software packages. The utility of the SP method is demonstrated in the analysis of non-metanic hydrocarbons (NMHC) concentration.  The analysis demonstrates that the prediction by the SP method captures the common pattern better than the existing prediction methods and also provides competitive prediction accuracy.\\

\noindent{\bf Key words}: Functional registration, Functional time series, (Spherical)$K$-means clustering, Nonlinear dimension reduction, Prediction, Shape space, State-space model.
\end{abstract}

\section{Introduction}

\begin{figure}[!h]
\centering
\includegraphics[scale=0.4]{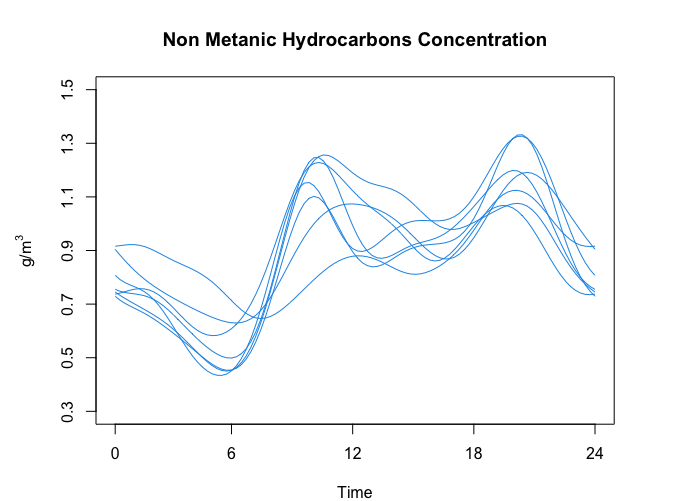}
\caption{7 consecutive NMHC concentration trajectories of one week}
\label{f1}
\end{figure}

When continuous-time records are separated into natural consecutive time intervals, such as days, weeks, or years, for which a reasonably similar behavior is expected, the resulting functions can be described as a functional time series. For functional time series data, one unit of observation is an observed trajectory. Functional time series data exhibits two types of variations: amplitude variation which corresponds to the size or scale of trajectory features, and phase variation which accounts for location variation and temporal shifts. In this paper, we analyzed curves of non-metanic hydrocarbon (NMHC) collected at road level in an Italian city. In Figure~\ref{f1}, 7 consecutive trajectories of NMHC concentration are displayed. Observe that each daily curve has two peaks. The precise timing of the peaks varies across days due to human activity or some other environmental reasons. The variation of the occurrence time of the peaks can be viewed as phase variation. However, existing work typically only consider the vertical variation (i.e., amplitude), but not the variation in phase. An immediate result is that, the predicted curve may not show the common underlying pattern. To overcome this serious limitation, we develop a novel method for stationary functional time series, whose trajectories share a common pattern. Our goal is not only to obtain competitive prediction from the past data by some stationary functional time series model, such as
functional auto-regression model, in terms of mean squared error, but also to preserve the underlying pattern for the predicted curves.

There are available prediction methods for stationary functional time series. Besse et al.~\cite{r4} proposed a non-parametric kernel predictor. Antoniadis and Sapatinas \cite{r1} studied first-order functional autoregression curve prediction based on a linear wavelet method. Kargin and Onaski \cite{r19} introduced the predictive factor method. Aue et al.~\cite{r3} developed a method that uses multivariate techniques in functional time series prediction. {Jiao et al.~\cite{r17} proposed a partial functional prediction method, for the cases where the functions to be predicted are partially observed.} There are also some other prediction methods for functional time series, and these methods have their own advantages. However, their main limitation is
that they incorporate only vertical variation among the curves. Hyndman and Shang \cite{r12} proposed to use weighted functional principal component regression and weighted functional partial least squares regression. One attractive feature of their method is its ability to take account for changes in the functional shapes over time. However, the geometrically decaying weights restrict the shape of each function to be close to the neighboring functions. Hence, while this method works well for processes with slowly-evolving functional shape, its disadvantage is that it might not be suitable for situations where ``neighboring" functions have different shapes or shapes change quickly across curves. Compared to Hydman's method, the SP method allows fast transition of phase. Some related work assume that functions are composed of multiple components which repeat themselves over different periods of time (see e.g.,~Lin et al.~\cite{r13} and Lin et al.~\cite{r14}). The difference between their work and our proposed method is that we assume that there is only one common pattern that repeats itself over the same period of time across curves. This, we believe, 
is more suitable for some cases such as the environmental data that is being 
analyzed in this paper.

When trajectories share a common pattern and meanwhile present phase variation, a typical technique researchers usually adopt is functional registration. In functional registration, each function is decomposed as $f_n(t)=X_n\circ\gamma_n(t)$, where the amplitude function $X_n(t)$ accounts for the vertical variation, and the warping function $\gamma_n(t)$ captures the phase information.  However, to the best of our knowledge, methods for functional time series prediction have not incorporated functional registration. The prediction method that we develop here involves the prediction of amplitude functions and warping functions. One of the major challenges is the prediction of warping functions, since they do not lie in a linear space, and thus ordinary linear models are not applicable. Warping functions must be monotonically increasing,  and are restricted to start and end at two fixed values. There are several ways to model warping functions. Generally speaking, all these methods seek to apply linear models to non-linear objects.

It is noted that warping functions share similar properties with probability distribution functions, that is, they are all non-decreasing and have common starting and ending values. There are some papers on modeling probability density functions. These work typically apply some transformations to density functions and then employ linear models to the transformed functions. Brumback and Lindstrom \cite{r5} proposed a self-modeling method for monotone functions involving the transformation proposed by Jupp \cite{r18}, which is a bijective map from the space of monotone increasing vectors to Euclidean space. Gervini \cite{r9} used the Jupp transformation to study warped functional regression. Peterson and M\"uller \cite{r23} proposed to use the log quantile density transformation and log hazard transformation to map a density function into a linear Hilbert space. Kokoszka et al.~\cite{r21} used the same transformations to predict density functions. Another way is to study the manifold structure of warping functions. Here some of these related methods are reviewed. Cheng and Wu \cite{r7} used local linear regression models to study the scale-to-manifold regression problem, where covariates lie on an unknown manifold. Su et al.~\cite{r28} employed the transported square-root vector field (TSRVF) to implement statistical analysis of trajectories on Riemannian manifolds. Dai and M\"uller \cite{r8} developed principal component analysis for sphere-valued functional data. They proposed to apply functional principal component analysis (fPCA) to the tangent vectors at the Fr\'echet mean of the sphere.

However, all these methods have some limitations. One common characteristic of the first kind of method is that the transformations all involve the ``logarithm'' which sometimes dictates the need of another re-scaling step {(e.g., $\log(f(Q(t)))$ and $\log(f(t)/\{1-F(t)\})$, where $F(t)$ and $Q(t)$ are the cumulative distribution function and quantile function of the density function $f(t)$, see Peterson and M\"uller \cite{r23}).} A major limitation of the logarithm function is that it will either shrink the variation of large values or exaggerate the variation of values close to 0. In addition, density functions (and warping functions) lie in a nonlinear space, and it is always unnatural to use linear models directly. {Regarding the second framework, one may consider applying linear models to the tangent space of the manifold composed of the square root of slope functions (SRSF) of warping functions $\gamma(t)$, defined as $\sqrt{\dot{\gamma}(t)}$. The SRSFs of warping functions lie on an infinite dimensional sphere, and thus the tangent space has clear and simple representation. However, the SRSFs of warping functions form only the positive orthant of the sphere ($-\sqrt{\dot{\gamma}(t)}$ is not included), and the predicted SRSFs by linear models may lie on the negative orthant.} In addition, this approach still seeks to transform nonlinear space to linear space, and thus also change the original variation. All of these problems motivate us to develop a new methodology to predict the stochastic process composed of warping functions.

We develop a novel method that can jointly predict the amplitude and warping
functions. The major advantage of our method is that it does not require any unnatural
transformations and it retains the predicted warping functions strictly in their original
non-linear space. To be more specific, we develop a state-space model where warping
functions are driven by hidden states, and consequently, there is no need to transform
warping functions between linear and non-linear spaces. We first implement functional
registration to obtain amplitude and warping functions. To predict warping functions,
we propose a state-space model, in which the states are driven by a Markov chain.
Spherical $K$-means clustering, which is a popular technique for dimension reduction of non-linear space, can reveal the potential low dimensionality of warping functions.
In the model, finite prototypes are employed to represent the nonlinear space of warping
functions, where each warping function is assumed to be the sum of its
corresponding prototype and a random error function. 

For the prediction of the amplitude function, we develop a varying coefficient operator functional auto-regression model. Varying-coefficient models have been extended to functional data. Sent\"urk and M\"uller \cite{r25} generalized the functional varying coefficient models to incorporate the influence of recent past values of predictors on current response. Further improvements were reported in Sent\"urk and M\"uller \cite{r26} which proposed a new representation for varying coefficient functions and introduced a smooth history index function to model the dependence of the response on the recent past values of predictors. Krafty et al.~\cite{r22} employed a varying coefficient model in the analysis of tumor growth curves. Our varying coefficient model is fully functional, and the states of previous warping functions influence the
current coefficient operators. The predicted warping functions and amplitude functions
are combined to obtain the final prediction.

In this article, the following issues will be addressed:
\begin{enumerate}
\item Since the real states in the state-space model are unknown in practice, the transition probability matrix of the hidden Markov chain has to be estimated through
the estimated states instead of the real states. The large-sample behavior of the
estimator will be investigated in this paper.
\item A method for determining the dimension and order of the varying coefficient
operator functional auto-regression model will be developed.
\item We will develop a measure to evaluate the performance of our proposed method in preserving the common pattern across curves.
\end{enumerate}

The rest of the paper is organized as follows. In Section \ref{s2}, we formulate the model for the stochastic process of warping functions and amplitude functions, the joint prediction procedure of amplitude and phase variation, and discuss how to measure shape similarity. In Section \ref{s3}, we derive the asymptotic properties of the least squares estimator of the transition matrix in the state-space model. Section \ref{s4} displays the results of the simulation study comparing the prediction performance of the SP method and some other competitor methods. In Section \ref{s5}, we report the results of the analysis on the NMHC concentration. Section \ref{s6} concludes the article.

\section{{Models, Algorithms, and Shape Similarity}}
\label{s2}
\subsection{Amplitude and phase variation}

In this section, we formulate the models for amplitude and phase variation. Let $\{f_n(t)\colon n\in\mathbb{N}\}$ be an arbitrary stationary functional time series defined on a common probability space $(\Omega,\mathcal{A},P)$, where the function index $n$ is discrete and the time index $t$ is continuous.  Assume the following decomposition $f_n(t)=X_n\circ\gamma_n(t).$ In this decomposition, $X_n(t)$ is the amplitude function and $\gamma_n(t)$ is the warping function. The observations $\{f_n(t)\colon n\in\mathbb{N}\}$ are elements of the Hilbert space $H=L^2[0,1]$ equipped with the inner product $\langle f_1,f_2\rangle=\int_0^1f_1(t)f_2(t)dt$, and $f_n(t)<\infty$ for any $t\in[0,1]$. The norm of each $f_n$ satisfies $\|f_n\|_2=\sqrt{\langle f_n,f_n \rangle}<\infty$. Define the mean function and covariance function of the amplitude functions as follows
$$\mu(t)=E\{X_n(t)\},\qquad K(t,s)=\text{cov}\{X_n(t), X_n(s)\}.$$
In practice, $\mu(t)$ and $K(t,s)$ are always unknown and need to be estimated from samples $\{X_1(t),\ldots,X_N(t)\}$ as follows:
$$\hat{\mu}(t)=\frac{1}{N}\sum_{n=1}^NX_n(t),\qquad \widehat{K}(t,s)=\frac{1}{N}\sum_{n=1}^N\{X_n(t)-\hat{\mu}(t)\}\{X_n(s)-\hat{\mu}(t)\}.$$
By Mercer's theorem, $K(t,s)$ and $\widehat{K}(t,s)$ admit the following decomposition
$$K(t,s)=\sum_{m=1}^\infty\lambda_m\nu_m(t)\nu_m(s),\qquad \widehat{K}(t,s)=\sum_{m=1}^\infty\hat{\lambda}_m\hat{\nu}_m(t)\hat{\nu}_m(s),$$
where $\langle\nu_{m_1},\nu_{m_2}\rangle=0$ ($m_1\ne m_2$) and $\|\nu_m\|_2=1$, $m\in\mathbb{N}_+$. The warping functions $\gamma_n\colon H\to H$ have the following property: $\gamma_n(0)=0$, $\gamma_n(1)=1$, $\gamma_n$ is invertible, both $\gamma_n$ and $\gamma_n^{-1}$ are continuous, {and assume that the first order derivative exists and satisfies $\dot{\gamma}_n(t)<\infty$ for all $t\in[0,1]$}. Let $\Gamma$ denote the set of all such functions. The square root of slope function (SRSF) of $\gamma_n(t)$ is defined as 
$$s_n(t)=S(\gamma_n(t))=\sqrt{\dot{\gamma}_n(t)},$$
and a SRSF $s_n(t)$ can be transformed back into a warping function $\gamma_n(t)$ by applying $S^{-1}(\cdot)$ to it
$$\gamma_n(t)=S^{-1}(s_n(t))=\int_0^ts_n^2(u)du,\qquad 0<t<1,$$
where $S(\cdot)$ is a bijective map. It can be shown that $\|s_n(t)\|=1$, thus $\{s_n(t)\colon n\in\mathbb{N}\}$ lie on an infinite-dimensional sphere. In practice, only $f_n$ is observed, and we propose to apply functional registration algorithm to obtain $X_n$ and $\gamma_n$. In the following, it is assumed that both $\{X_n(t)\colon n\in\mathbb{N}\}$ and $\{\gamma_n(t)\colon n\in\mathbb{N}\}$ are already obtained by functional registration (see e.g.,~Ramsay and Silverman \cite{r29}, Kneip \& Ramsay \cite{r20}, and Srivastava \& Klassen \cite{r27}).

\subsection{Functional auto-regression model for amplitude functions}
One way to model amplitude functions is by a FAR model. A FAR($q$) process is defined by the stochastic recursion
$$X_n-\mu=\sum_{j=1}^q\Phi_j(X_{n-j}-\mu)+\epsilon_n,$$
where $\{\epsilon_n(t)\colon n\in\mathbb{Z}\}$ are centered, independent and identically distributed 
innovations in $L^2[0,1]$ and  $\Phi_j(\cdot) \colon H \to H$ is a bounded linear operator for $j=1,\ldots,q$, {and are defined so that the above recursive equation has a unique causal solution (see \cite{r11}, pp.~236). Horv\'ath and Kokoszka \cite{r11} has developed a sufficient condition for causality of FAR(1) process, and the result can be extended to FAR($q$) process ($q>1$) with the state-space form of FAR($q$) process.

The FAR model is easy-to-implement for the prediction of functional time series. One approach of estimation is to first project functions onto a finite dimensional sub-space spanned by some functional basis, e.g.,\ functional principal components (fPC), then multivariate techniques can be applied without much loss of information (see Aue et al.~\cite{r3}). 
\subsection{State-space model for warping functions}
\label{s2.3}
{Phase variation, which pertains to the variation of locations of curve features, is captured by the warping functions $\{\gamma_n(t)\colon n\in\mathbb{N}\}$}. Since $\{\gamma_n(t)\colon n\in\mathbb{N}\}$ are defined in an infinite dimensional non-linear {manifold}, linear methods are not appropriate for the prediction of $\{\gamma_n(t)\colon n\in\mathbb{N}\}$.  Note that it is computationally intractable to predict warping functions in the infinite dimensional {manifold} $\Gamma$. {Hence, we propse to 
employ non-linear dimensional reduction techniques, and develop the state-space model with the following assumptions.}
\begin{itemize}
\item The process $\{\gamma_n(t)\colon n\in\mathbb{N}\}$ is driven by a Markov chain, which is irreducible, ergodic, and of finite states. Each state $c_n$ of the Markov chain is associated with a fixed prototype warping function $b_{c_n}(t)$. $\gamma_n(t)$ can be expressed as the summation of one prototype and a random error function $u_n(t)$. 
\item The random error functions $\{u_n(t)\colon n\in\mathbb{N}\}$ are of mean zero, and given $c_n$, $u_n(t)$ is independent of $c_m$ and $u_{m}(t)$, $m\ne n$, and are such that the resulting function $\gamma_n(t)$ is still a warping function. 
\end{itemize}
Suppose the Markov chain has $g$ states, then each state $c_n$ can be represented by a state-indicating row vector $\omega_n$, which is $g$-dimensional satisfying $\omega_{n,c_n}=1$ and $\omega_{n,i}=0$, for $i\ne c_n$. Denote $P$ as the transition probability matrix. The state-space model is specified as follows
\begin{align*}
E[\omega_{n}|\omega_1,\ldots,\omega_{n-1}]&=E[\omega_{n}|\omega_{n-1}]=\omega_{n-1}P,\\
\gamma_n(t)&=\sum_j^g\omega_{n,j}b_j(t)+u_n(t).
\end{align*}
 The second equation is an analogue of the fPC representation of functions in $L^2[0,1]$. The prototypes $\{b_j(t)\colon j=1,\ldots,g\}$ can be viewed as a series of basis functions of $\Gamma$, and the state-space model is a discrete approximation of the continuous evolution of warping functions. 
The number of prototypes depends on the variability of warping functions. Under high variability, a large number of prototypes are typically needed to approximate true dynamic process $\{\gamma_n(t)\colon n\in\mathbb{N}\}$ well by the state-space process.

\begin{remark}In this state-space model, warping functions are driven by hidden states and it is not needed to employ linear models to account for its functional variability, and consequently, there is no need to transform warping functions between linear and non-linear spaces. 
\end{remark}

\begin{remark}One possible concern is identifiability, say, $X\circ\gamma_1$ and $X\circ(\gamma_0^{-1}\circ\gamma_0)\circ\gamma_1$ are the same, where $\gamma_0$ and $\gamma_1$ are two arbitrary different warping functions. However, once the template is fixed (e.g.,~sample mean) for functional registration, it will not be a problem. Specifically, if the template $X_0(t)$ is fixed for $\{f_n\colon n\in\mathbb{N}\}$, the warping function $\gamma_n=\arg\min_{\gamma\in\Gamma}\|f_n-X_0\circ\gamma\|$ is unique for arbitrary $n$, where $\|\cdot\|$ is some metric employed for functional registration, and in this paper, we propose to use Fisher-Rao metric which will be discussed later.
\end{remark}

\subsubsection{Estimation of the state-space model}
Since the hidden states and transition probability matrix are unknown in practice, we need to first estimate $b_j$'s, $\omega_n$'s, and then $P$. We apply spherical $K$-means clustering, which is a widely-accepted dimension reduction technique for non-linear space, to the SRSFs of warping functions, and use the cluster centroids as the estimators of the SRSFs of $b_j$'s. The estimators of $b_j$'s can be obtained by applying $S^{-1}(\cdot)$ to the cluster centroids, 
$$\hat{b}_j(t)=S^{-1}(\hat{p}_j(t)),\qquad j=1,\ldots,g,$$
where $\hat{p}_j(t)$ is the centroid of the $j$-th cluster of SRSFs. The classified categories of $\{s_n(t)\colon n\in \mathbb{N}\}$ are considered as the estimated states of $\{\gamma_n(t)\colon n\in \mathbb{N}\}$. More details are discussed below.

The standard spherical $K$-means clustering aims to minimize
$$D=\sum_{n=1}^N(1-\cos(s_n,p_{c_n}))=\sum_{n=1}^N(1-\langle s_n,p_{c_n}\rangle)$$
over all assignments of objects $n$ to cluster $c_n\in\{1,\ldots,g\}$ and over all SRSF representations of prototype warping functions $p_1,\ldots,p_g$. A typical projection and minimization procedure is repeated to obtain $\hat{c}_n$'s and $\hat{p}_j$'s. 

Let $\hat{\omega}_n$ denote a $g$-dimensional vector where only the $\hat{c}_n$-th element is $1$ and the rest elements are zeros. Then $P$ is estimated by the least squares method, where $\omega_n$ is replaced with $\hat{\omega}_n$, say, $$\widehat{P}=\arg\min\limits_{P}\sum_{n=2}^{N}\|\hat{\omega}_{n}-\hat{\omega}_{n-1}P\|^2_2.$$

The number of hidden states is unknown in practice, and we propose a cross-validation method in Section~\ref{stateselection} to select $g$. We assume the selected $g$ is correct, and will not distinguish between the selected $g$ and the real number of states. Note that, using the R package {\bf skmeans}, spherical $K$-means clustering algorithm can be implemented by the R function $skmeans$ (see Hornik et al.~\cite{r10}). The estimation procedure is summarized in Algorithm~\ref{al1}:
\begin{center}
\begin{minipage}{\dimexpr\linewidth-1em}
\begin{algorithm}[H]
\caption{Estimation of the state-space model}
\label{al1}
\justifying\hspace{-1.4em}\textbf{Step 1} Obtain the SRSFs of warping functions, $s_n=S(\gamma_n)$.\\
\textbf{Step 2} Fix the number of states $g$, apply spherical $K$-means clustering to $\{s_n\colon n\in\mathbb{N}\}$, and obtain the cluster centroids $\{\hat{p}_j\colon j=1,\ldots,g\}$ and the classified categories $\{\hat{c}_n\colon n\in \mathbb{N}\}$.\\
\textbf{Step 3} Apply $S^{-1}(\cdot)$ to $\{\hat{p}_j\colon j=1,\ldots,g\}$ to obtain the estimated prototype warping functions, say, $\hat{b}_j=S^{-1}(\hat{p}_j),\ j=1,\ldots,g.$
\end{algorithm}
 \end{minipage}
 \end{center}
\subsection{Joint Prediction Methodology}
After separating amplitude and phase components, it is natural to consider how to predict the two components jointly, as they are not necessarily independent of each other. Because warping functions and amplitude functions are defined in two different spaces, it is necessary to find a common space for these two kinds of functions in the joint prediction. To be more specific, we assume the amplitude and warping functions are jointly driven by a Markov process. The joint modeling procedure is discussed below.

\subsubsection{Prediction of warping function}
We convert the stochastic process of warping functions into a Markov chain by applying spherical $K$-means clustering to their corresponding SRSFs, as has been discussed in Section~\ref{s2.3}. In order to incorporate the correlation between phase and amplitude variation, we assume the same kind of state-space model for amplitude functions, and apply $K$-means clustering to estimate the hidden states of amplitude functions. Similarly, the classified categories are treated as the estimated hidden states. Figure~\ref{f2} shows the framework, where $\omega$ represents the true state and $\hat{\omega}$ represents the estimated state, and superscripts $(a)$ and $(f)$ refer to amplitude and phase variation respectively.

\begin{figure}[!h]
\centering
\includegraphics[scale=0.45]{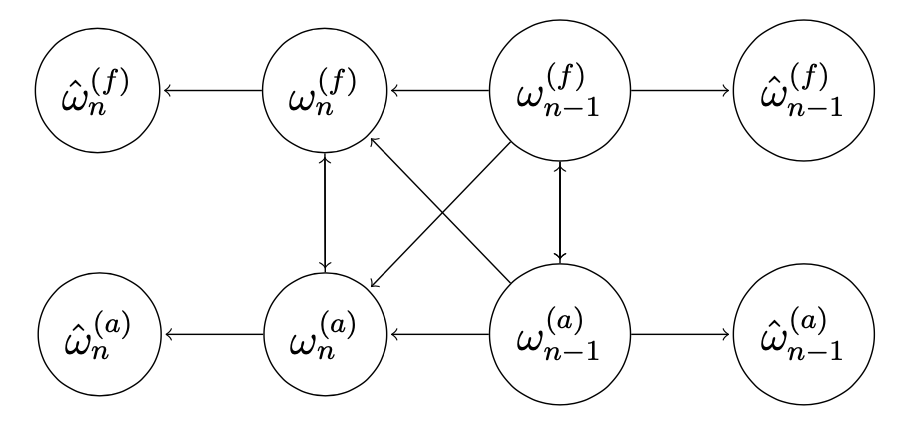}
\caption{Hidden states and estimated states.}
\label{f2}
\end{figure}

The two categorical sequences are combined to obtain a new sequence, $\hat{\omega}_{n}=(\hat{\omega}_n^{(f)}\otimes\hat{\omega}_n^{(a)})$, where $\otimes$ signifies the Kronecker product. Then apply the least squares method to estimate the transition matrix ${P}$ of this combined estimated Markov chain, where $P$ is a $g\ell\times g\ell$ matrix, $g$ is the number of states of phase variation and $\ell$ is the number of states of amplitude variation. Then the predicted state is
$\hat{\hat{\omega}}_{N+1}=\hat{\omega}_N\widehat{{P}}.$
Suppose $\hat{\hat{\omega}}_{N+1}^{(f)}$ is the predictor of ${\omega}^{(f)}_{N+1}$, which is obtained from $\hat{\hat{\omega}}_{N+1}$ as $\hat{\hat{\omega}}^{(f)}_{N+1}=\hat{\omega}_N\widehat{P}J,$
where $J$ is the $(g\ell)\times g$ matrix
\begin{equation*}
J=\left(\begin{array}{cccc}
{\bf 1}_\ell&{\bf 0}_\ell&\cdots&{\bf 0}_\ell\\
{\bf 0}_\ell&{\bf 1}_\ell&\cdots&{\bf 0}_\ell\\
\vdots&\vdots&\ddots&\vdots\\
{\bf 0}_\ell&{\bf 0}_\ell&\cdots&{\bf 1}_\ell\\
\end{array}\right),
\end{equation*}
and ${\bf 1}_\ell=(1,\ldots,1)'_{1\times\ell}$, ${\bf 0}_\ell=(0,\ldots,0)'_{1\times\ell}$, and $'$ signifies transposition. The predicted warping function is 
$\hat{\gamma}_{N+1}(t)=\sum\limits_{j=1}^g\hat{\hat{\omega}}^{(f)}_{N+1,j}\hat{b}_j(t).$
As a side note, $\hat{\hat{\omega}}_{N+1}$ is not a state-indicating vector, but a vector of probabilities of which the sum is one.
It can be easily shown that $\hat{\gamma}_{N+1}(t)$ is a warping function since 
\begin{align*}
&\frac{d}{dt}\hat{\gamma}_{N+1}(t)=\sum\limits_{j=1}^g\hat{\hat{\omega}}^{(f)}_{N+1,j}\frac{d\hat{b}_j(t)}{dt}\ge0,\\
&\sum_{j=1}^g\hat{\hat{\omega}}^{(f)}_{N+1,j}\hat{b}_j(1)=\sum_{j=1}^g\hat{\hat{\omega}}^{(f)}_{N+1,j}=1,\qquad \sum_{j=1}^g\hat{\hat{\omega}}^{(f)}_{N+1,j}\hat{b}_j(0)=0.
\end{align*}
\begin{remark}
When the sample size is small, some ad-hoc adjustments might be needed to let $\widehat{P}$ satisfy the constraints of a transition matrix. One approach is to obtain the transition matrix by solving the optimization problem  
$\widehat{P}=\arg\min\limits_{P\in P_\mathcal{M}}\|P-\widehat{{P}}^{\rm LS}\|_F,$
where $P_\mathcal{M}$ is the set of all probability transition matrices, $\|\cdot\|_F$ is Frobenius norm, and $\widehat{P}^{\rm LS}$ is the original least squares estimator of $P$. 
\end{remark}
\subsubsection{Data-driven selection of the number of states}
\label{stateselection}
To the best of our knowledge, there is no widely accepted procedure for order selection of hidden Markov models. The selection of state numbers is a trade-off between bias and variance. A large number of states will decrease the approximation error by prototype warping functions, but will increase variance since there will be more parameters to be estimated. Considering that our purpose is prediction, we propose an approach based on prediction error. The prediction performance is evaluated by $\ell^2$ mean squared error and amplitude distance (Section \ref{ampl}). Assume that there is a large test data-set $D_{\rm{test}}$ which is an independent copy of the dataset used for model fitting.  Then use the first $80\%$ curves in $D_{\rm{test}}$ to fit a model with {$g$ phase states and $\ell$ amplitude states}, and predict the rest $20\%$ curves with the fitted model, and then calculate the mean squared error and the average amplitude distance between the predicted curves and the curves to be predicted. Then refer to these two errors for the order selection.
In practice, the sample size may be limited, and it is not possible to reserve a large fraction
of data for the testing set. In this case, Monte-Carlo cross-validation is a good alternative approach. A fraction of consecutive curves are selected as training set
and the rest curves are used for testing. This procedure is repeated multiple times where
the partitions are randomly chosen on each run. A group of candidate state
numbers are preset, and the two average errors are computed for models with different candidates. The state numbers are selected such that both errors are decent. 

\subsubsection{Prediction of amplitude function}
\label{s2.4.2}
We now develop the FAR model with varying coefficient operators for the prediction of amplitude functions. The coefficient operator is determined by the state of the previous warping function. 
Define $Y_{n}(t)=X_n(t)-\mu(t)$ and let $c_n^{(f)}$ be the hidden state of $\gamma_n$. The proposed model has the following representation
$$Y_{n+1}=\sum_{h=1}^q\Phi^{(c_n^{(f)})}_{h}(Y_{n+1-h})+\epsilon_{n+1},$$
where $\{\epsilon_n\colon n\in\mathbb{N}\}$ are centered, independent and indentically distributed innovations in $L^2[0,1]$, and $\{\Phi^{(k)}_{h}\colon k=1,\ldots,g, h=1,\ldots,q\}$ are bounded linear operators, {and are constrained so that the above recursive equation has a unique causal solution.}

The estimation of $\{\Phi^{(k)}_{h}\colon k=1,\ldots,g, h=1,\ldots,q\}$ is obtained by minimizing the objective function
$$S(\Phi)=\sum_{n=h}^{N-1}\left\|Y_{n+1}-\sum_{h=1}^q\Phi^{{(c}_{n}^{(f)})}_{h}(Y_{n+1-h})\right\|^2_2.$$
By simple decomposition, 
$$S(\Phi)=\sum_{k=1}^g\sum_{n_k=1}^{N_k}\left\|Y_{n_k+1}-\sum_{h=1}^q\Phi_{h}^{(k)}(Y_{n_k+1-h})\right\|^2_2,$$
where $N_k$ is the number of $Y_{n+1}$ {so that $\gamma_n$ is in state $k$}. Then minimize the following quantity to obtain the estimation of $\{\Phi^{(k)}_{h}\}_{h=1}^p$:
$$S_k(\Phi)=\sum_{n_k=1}^{N_k}\left\|Y_{n_k+1}-\sum_{h=1}^q\Phi_{h}^{(k)}(Y_{n_k+1-h})\right\|^2_2.$$
After projecting all functional elements onto the sub-eigenspace spanned by the finite major functional principal components of $\{X_n(t)\colon n\in\mathbb{N}_+\}$, the multivariate technique can be applied to estimate $\{\Phi^{(k)}_h\colon h=1,\ldots,q\}$ for each $k$. 
Denote $\widehat{\Phi}_h^{(k)}$ as the estimator of ${\Phi}_h^{(k)}$, then the predictor of $Y_{N+1}$ is $$\widehat{Y}_{N+1}=\sum_{k=1}^g\sum_{h=1}^{q}\widehat{\Phi}_h^{(k)}(Y_{N+1-h})\mathbbm{1}(\hat{c}_N=k).$$
The entire joint prediction procedure is summarized in Algorithm~\ref{al2}.
\begin{center}
\begin{minipage}{\dimexpr\linewidth-1em}
\begin{algorithm}[H]
\caption{Joint prediction algorithm (one-step ahead)}
\label{al2}
\justifying\hspace{-1.4em}\textbf{Step 1} Apply functional registration algorithm to obtain the amplitude and warping functions.\\
\textbf{Step 2} Apply spherical $K$-means clustering algorithm resp.~$K$-means clustering algorithm to the SRSFs of the warping functions resp.~the amplitude functions to obtain the estimated states. Combine the two state sequences, fit a Markov model, and obtain the prediction of  the next warping function $\hat{\gamma}_{N+1}$ by the state space model.\\
\textbf{Step 3} Obtain the prediction of the next amplitude function, $\widehat{Y}_{N+1}$, based on a FAR model with varying coefficient operators. \\
\textbf{Step 4} Warp $\widehat{Y}_{N+1}+\hat{\mu}$ by $\hat{\gamma}_{N+1}$ to obtain the final prediction, $\hat{f}_{N+1}=(\widehat{Y}_{N+1}+\hat{\mu})\circ\hat{\gamma}_{N+1}.$
\end{algorithm}
 \end{minipage}
 \end{center}
\begin{remark} 
The final expression is binary. In practice, the weighted predictor can also be considered,
$\widehat{Y}_{N+1}=\sum\limits_{k=1}^g\sum\limits_{h=1}^{q}\widehat{\Phi}_h^{(k)}(Y_{N+1-h})P(\hat{c}_N=k).$
The weighted predictor typically have smaller variance but larger bias. The probabilities of states $P(\hat{c}_N=k)$ need to be estimated under some model, for example, $P(\hat{c}_N=k)\propto1/d(\hat{\gamma}_N,\hat{b}_k)$, where $d(\hat{\gamma}_N,\hat{b}_k)=1-\cos(S(\hat{\gamma}_N),S(\hat{b}_k))$. 
\end{remark}

\subsubsection{Parameter selection}
Now we develop the functional final prediction error (fFPE) criterion to select the order and dimension of the sub-eigenspace for the prediction of amplitude functions. Create the $d$-variate fPC score vector $\bm{Y}_n=(y_{n,1},\ldots,y_{n,d})'$, 
where $y_{n,m}=\langle Y_n,\nu_m\rangle=\langle X_n-\mu,\nu_m\rangle$. Since the eigenfunctions are orthogonal and the fPC scores are uncorrelated for each $Y_n$, the mean squared prediction error can be decomposed as
\allowdisplaybreaks
\begin{align*}
E\left[\left\|Y_{N+1}-\widehat{Y}_{N+1}\right\|^2\right]&=E\left[\left\|\sum_{m=1}^\infty y_{N+1,m}\nu_m-\sum_{m=1}^d\hat{y}_{N+1,m}\nu_m\right\|^2\right]\\
&=E\left[\left\|\bm{Y}_{N+1}-\widehat{\bm{Y}}_{N+1}\right\|^2\right]+\sum_{m>d}\lambda_m,
\end{align*}
where $\|\cdot\|$ denotes the $\ell^2$-norm, and $\hat{y}_{N+1,m}$ is the prediction of $y_{N+1,m}$ from the past $d$-variate fPC score vectors. As for the first summand, assume $\{\bm{Y}_n\colon n\in\mathbb{N}\}$ follows a $d$-variate VAR($q$) process (see Aue et al.~\cite{r3} for the justification of the VAR process) with varying coefficient matrix, that is,
$\bm{Y}_{n+1}=\bm{\Phi}_1^{(c_n^{(f)})}\bm{Y}_n+\ldots+\bm{\Phi}_q^{(c_n^{(f)})}\bm{Y}_{n-q+1}+\bm{ Z}_{n+1},$
where $\bm{ Z}_n$ is the error term. For any state of warping function $k$, it can be shown that (see, e.g.,\ L\"utkepohl \cite{r15})
$\sqrt{N_k}(\hat{\bm{\beta}}_k-\bm{\beta}_k)\overset{\mathcal{L}}\to\mathcal{N}_{qd^2}(0,\Sigma_{Z,k}^d\otimes\Gamma_{q,k}^{-1}),$
where $\bm{\beta}_k=\text{vec}([\bm{\Phi}_1^{(k)},\ldots,\bm{\Phi}_q^{(k)}]')$ and $\hat{\bm{\beta}}_k$ is the least squares estimator of $\bm{\beta}_k$, $\Sigma_{Z,k}^d$ is the covariance matrix of $\{\bm{Z}_{n+1}\colon n\in\mathbb{N}\}$ and $\Gamma_{q,k}=\text{var}(\text{vec}([\bm{Y}_q,\ldots,\bm{Y}_1]))$ as $c_n^{(f)}=k$.
Let $\widehat{\bm{Y}}^{(k)}_{N+1}$ be the predictor of $\bm{Y}_{N+1}$ as $c_N^{(f)}=k$. Assuming the classification is correct, it follows that
\allowdisplaybreaks
\begin{align*}
&E\left[\left\|\bm{Y}_{N+1}-\widehat{\bm{Y}}_{N+1}\right\|^2\right]=E\left[\left\|\bm{Y}_{N+1}-\sum_{k=1}^g\widehat{\bm{Y}}^{(k)}_{N+1}\mathbbm{1}({c}_N^{(f)}=k)\right\|^2\right]\\
&=E\left[E\left[\left\|\bm{Y}_{N+1}-\widehat{\bm{Y}}_{N+1}^{({c}_N^{(f)})}\right\|^2\Bigg|{c}_N^{(f)}\right]\right]=\sum_{k=1}^gE\left[\left\|\bm{Y}_{N+1}-\widehat{\bm{Y}}^{(k)}_{N+1}\right\|^2\right]P({c}_N^{(f)}=k)\\
&=\sum_{k=1}^gE\left[\left\|\bm{Y}_{N+1}-\sum_{h=1}^q\widehat{\bm{\Phi}}^{(k)}_{h}\bm{Y}_{N+1-h}\right\|^2\right]P({c}_N^{(f)}=k)\\
&=\sum_{k=1}^g\left\{\mbox{tr}(\Sigma^d_{Z,k})+E\left[\left\|\sum_{h=1}^q(\bm{\Phi}^{(k)}_{h}-\widehat{\bm{\Phi}}^{(k)}_{h})\bm{Y}_{N+1-h}\right\|^2\right]\right\}P({c}_N^{(f)}=k)\\
&=\sum_{k=1}^g\left\{\mbox{tr}(\Sigma^d_{Z,k})+\sum_{k=1}^gE\left[\left\|I_p\otimes(\bm{Y}'_N,\ldots,\bm{Y}'_{N-q+1})(\bm{\beta}_k-\hat{\bm{\beta}}_k)\right\|^2\right]\right\}P({c}_N^{(f)}=k)\\
&\sim \sum_{k=1}^g\left\{\text{tr}(\Sigma^d_Z)+\frac{qd}{N_k}\text{tr}(\Sigma^d_Z)\right\}P({c}_N^{(f)}=k),
\end{align*}
where $a_N\sim b_N$ means $a_N/b_N\to 1$. Finally we conclude that
$$E[\|Y_{N+1}-\widehat{Y}_{N+1}\|^2]\sim\sum_{k=1}^g\left(\frac{N_k+qd}{N_k}\right)\ \text{tr}({\Sigma}^d_{Z,k})P(c_N^{(f)}=k)+\sum_{m>d}{\lambda}_m.$$
Replacing $\text{tr}(\Sigma^d_{Z,k})$ with $\text{tr}(\widehat{\Sigma}^d_{Z,k})$, $P({c}_N^{(f)}=k)$ with $N_k/N$, and $\lambda_m$ with $\hat{\lambda}_m$, where $\widehat{\Sigma}_{Z,k}^d$ is the unbiased estimator of $\Sigma_{Z,k}^d$, 
the fFPE criterion is given by, 
$$\text{fFPE}(q,d)=\sum_{k=1}^g\left(\frac{N_k+qd}{N}\right) \text{tr}(\widehat{\Sigma}^d_{Z,k})+\sum_{m>d}\hat{\lambda}_m.$$
We propose to select $q$ and $d$ by minimizing fFPE($q,d$).

\subsection{Shape Similarity}

\subsubsection{Functional shape space}
One of the main questions considered in this article is: what is a good  measurement of shape similarity? In order to compare the shapes of different trajectories, we need to formally define the functional shape space $\mathcal{E}$ and to evaluate shape similarity. Here, we shall follow the convention that shape is independent of scale and location. We first rescale and relocate functions, so that they are of unit norm, and start at the same value. Then we study the shape difference of the thus obtained set. This resulting space is termed pre-shape space. 

Suppose there are two functions $f_1$ and $f_2$, with {the corresponding transformations} in the pre-shape space as $\tilde{f}_1$ and $\tilde{f}_2$. We propose the principle that, if $\tilde{f}_1$ can be warped into $\tilde{f}_2$, the two functions $f_1$ and $f_2$ are considered to be of the same shape. This idea is motivated by shape data analysis (see e.g.,~Srivastava and Klassen \cite{r27}). To be specific, stretching, rotating, or relocating do not change the shape of planar shape objects. As a motivating example in shape data analysis, suppose that there is a planar contour delineating a human hand, stretching, rotating, and relocating the contour will not change the shape of human hand. 

In the functional shape space, we unify the shape representations, that is, obtain the unification of all points in pre-shape space representing the same shape. 
Therefore, the functional shape space $\mathcal{E}$ is defined as the quotient space of $L^2[0,1]$ with respect to relocating, rescaling and warping. We define the equivalence relation $\equiv$ on $\mathcal{E}$ as follows: let $\tilde{f}_1$, $\tilde{f}_2$ be the pre-shape elements of two functions $f_1$, $f_2$, then ${f}_1\equiv {f}_2$ if there exists a warping function $\gamma$ such that $\tilde{f}_1=\tilde{f}_2\circ\gamma$.
For any function $f_0$, the set of all functions, of which transformations in the pre-shape space can be warped into $\tilde{f}_0$, is considered as an object in the functional shape space $\mathcal{E}$, that is,
$[f_0]=\{f\colon \tilde{f}\circ\gamma=\tilde{f_0},\gamma\in\Gamma\}\in\mathcal{E}.$
Based on this definition, the distance $d([f_1],[f_2])$ between two shape objects $[f_1]$ and $[f_2]$ should be invariant to relocating, rescaling and warping of $f_1$ and $f_2$. 
\subsubsection{Amplitude distance}
\label{ampl}
For any $f\in {H}_0=\{f\in {H}\colon\dot{f}>0\}$, and $\nu_1,\nu_2\in T_f({H})$, where $T_f({H})$ is the tangent space of $H$ at $f$, {defined as $\{h\in H\colon \langle f,h\rangle=0\}$}, the Fisher--Rao metric is defined as the inner product
$$\langle\langle\nu_1,\nu_2\rangle\rangle_f=\frac{1}{4}\int_0^1\dot{\nu}_1(t)\dot{\nu}_2(t)\frac{1}{\dot{f}(t)}dt.$$
One important property of Fisher-Rao metric is invariance of simultaneous warping: for any $\gamma\in\Gamma$, $d_{\rm{FR}}(f_1,f_2)=d_{\rm{FR}}(f_1\circ\gamma,f_2\circ\gamma)$, where $d_{\rm{FR}}$ denotes the geodesic distance induced by the Fisher-Rao metric. 
Under the SRSF representation, the Fisher--Rao Riemannian metric on $H_0$ becomes the standard $\ell^2$-metric (see \cite{r27}, pp.~106). With this property, the geodesic distance under the Fisher--Rao metric can be written explicitly as
$d_{\rm FR}(f_1,f_2)=\|s_1-s_2\|_2,$
where  $s_1,s_2$ are the SRSF representations of $f_1,f_2$. 
The Fisher--Rao metric is defined only on a subset $H_0\subset H$, but under SRSF representation, it can be generalized to $H$ endowed with the $\ell^2$-metric. The $\ell^2$-metric on SRSF representation space are the extended Fisher--Rao metric.

We shall use the amplitude distance~(\ref{amdis}), which has been shown to be a proper distance in the functional shape space, to measure shape similarity, 
\begin{equation}
\label{amdis}
d([f_1],[f_2])=\inf\limits_{\gamma}d_{\rm{FR}}(\tilde{f}_1,\tilde{f}_2\circ\gamma).
\end{equation}
If two functions are of the same shape, then the amplitude distance between the two functions is zero. The geodesic distance under the Fisher--Rao metric is invariant to simultaneous warpings. Therefore, the effect of phase variation does not influence the amplitude distance between two functions, say, {for any two different warping functions $\gamma_1$ and $\gamma_2$},
$\inf_{\gamma}d_{\rm{FR}}(\tilde{f}_1\circ\gamma_1,\tilde{f}_2\circ\gamma_2\circ\gamma)=\inf_{\gamma}d_{\rm{FR}}(\tilde{f}_1,\tilde{f}_2\circ\gamma),$
and thus the amplitude distance between two shape objects is unique. (see \cite{r27}, pp.~85--88)

\begin{remark}
In this paper, we use both the amplitude distance and the Euclidean distance to evaluate the prediction. Neither of the distance can evaluate the prediction individually, as we consider both amplitude and phase variation. 
\end{remark}

\section{Theoretical Results}
\label{s3}
The least squares method is employed to estimate the unknown transition probabilities, and aim to find the asymptotic properties of the estimator. It is known that the least squares estimator of the transition matrix of a Markov chain is consistent and asymptotically normal (see van der Plas \cite{r30}). However, since the real hidden states need to be estimated, the least squares estimator of the transition matrix ${P}$ is not necessarily consistent with ${P}$. To find the matrix that $\widehat{P}$ is consistent with, the following assumptions are needed.
\begin{enumerate}
\item[A1.] The Markov chain $\{\omega_n\colon n\in\mathbb{N}\}$ is stationary and ergodic, and has finite states;
\item[A2.] The estimated prototypes are obtained from an independent copy of observations, and thus the estimated state $\hat{\omega}_n^{(a)}$ resp.~${\hat{\omega}}_n^{(f)}$ is independent of $\mathcal{F}_{a,0}^{\infty}$ and $\mathcal{F}^{\infty}_{f,0}$ given $\omega_n^{(a)}$ resp.~$\omega_n^{(f)}$, where 
$\mathcal{F}_{a,0}^\infty=\sigma(\omega^{(a)}_0,\hat{\omega}^{(a)}_0,\ldots,\omega^{(a)}_\infty,\hat{\omega}^{(a)}_\infty)$ and $\mathcal{F}_{f,0}^\infty=\sigma(\omega^{(f)}_0,\hat{\omega}^{(f)}_0,\ldots,\omega^{(f)}_\infty,\hat{\omega}^{(f)}_\infty)$, and $\sigma(X)$ signifies the $\sigma$-algebra induced by $X$;
\item[A3.] $p(\hat{c}^{(f)}_n=\beta|c^{(f)}_n=\alpha)$ are the same across $n$ for any $\alpha,\beta=1,\ldots,g$, and $p(\hat{c}^{(a)}_n=\beta'|c^{(a)}_n=\alpha')$ are the same across $n$ for any $\alpha',\beta'=1,\ldots,\ell$. 

\end{enumerate}
Note that Assumption (A2) is compatible with the assumption on the error term $u_n$ of the state-space model. Based on the model assumption, the estimated state $\hat{\omega}_n$ is only relevant to the real state $\omega_n$ and the random error $u_n$, so Assumption (A2) is a natural consequence of the assumption on $u_n$. Assumption (A2) means, given the corresponding real state, the estimated state is independent of all other states. This is a reasonable assumption, since as the sample size grows large enough, the estimated prototype functions tend to be uncorrelated with any individual function. Assumption (A3) guarantees a constant transition probability matrix of the estimate states.

The Bayesian theorem implies the following proposition.
\begin{prop}
\label{prop1}
Under Assumptions (A1)---(A3), the transition probabilities of the combined estimated process $\{\hat{\omega}^{(f)}_n\otimes\hat{\omega}^{(a)}_n\colon n\in\mathbb{N}\}$ are given by
\begingroup
\allowdisplaybreaks
\begin{align*}
P(\hat{\omega}^{(f)}_{n+1},\hat{\omega}^{(a)}_{n+1}|\hat{\omega}^{(f)}_{n},\hat{\omega}^{(a)}_{n})&=\\
&\hspace{-3cm}\sum_{\omega_{n+1}^{(f)},\omega_{n+1}^{(a)},\omega_n^{(f)},\omega_n^{(a)}} P(\omega^{(f)}_{n+1},{\omega}^{(a)}_{n+1}|\omega^{(a)}_{n},\omega^{(f)}_{n})P(\hat{\omega}^{(a)}_{n+1}|{\omega}^{(a)}_{n+1})P(\hat{\omega}^{(f)}_{n+1}|{\omega}^{(f)}_{n+1})\\
&\hspace{-0.5cm}\times \frac{P(\hat{\omega}^{(a)}_{n}|{\omega}^{(a)}_{n})P(\hat{\omega}^{(f)}_{n}|{\omega}^{(f)}_{n})P(\omega^{(a)}_{n},\omega^{(f)}_{n})}{\sum_{\omega_n^{(a)},\omega_n^{(f)}}P(\hat{\omega}^{(a)}_{n}|{\omega}^{(a)}_{n})P(\hat{\omega}^{(f)}_{n}|{\omega}^{(f)}_{n})P({\omega}^{(f)}_{n},{\omega}^{(a)}_{n})}.
\end{align*}
\endgroup
\end{prop}
\begin{remark}Proposition \ref{prop1} implies the transition probability of the estimated Markov chain. 
\end{remark}
Next we show that the least squares estimator $\widehat{P}$ is consistent with $\widetilde{P}=\{P(\hat{\omega}^{(f)}_{n+1},\hat{\omega}^{(a)}_{n+1}|\hat{\omega}^{(f)}_{n},\hat{\omega}^{(a)}_{n})\}$.
Notationally, let $L_N(P)=N^{-1}\sum\limits_{n=2}^N\|\hat{\omega}_{n}-\hat{\omega}_{n-1} P\|^2,$
then we develop the following theorem for the least squares estimator $\widehat{P}_N$, which is a generalization of the result of van der Plas \cite{r30}.
\begin{thm}
\label{consis}
Under Assumptions (A1)---(A3), for each $N$ there exists a random matrix $\widehat{P}_N$ such that $L_N(\widehat{P}_N)=\inf\limits_{P}L_N(P)$ and
$\lim\limits_{N\to\infty}\widehat{P}_N=\widetilde{P} \ \  a.s..$
\end{thm}
In order to establish the asymptotic normality of the least squares estimator $\widehat{P}_N$, we make one additional assumption as follows,
\begin{enumerate}
\item[A4.] The matrix $A=\{a_{ij}\}\ \text{where} \ a_{ij}=2E\{\langle\frac{\partial\hat{\omega}_0{P}}{\partial \theta_i}\vert_{\widetilde{P}},\frac{\partial\hat{\omega}_0{P}}{\partial \theta_j}\vert_{\widetilde{P}}\rangle\}$ is positive definite, where $\theta_i$ is the $i$-th element of $\mbox{vec}({P})$.
\end{enumerate} 
and introduce the following notations,
\begin{align*}
F_i(n,\theta)&=\left\langle\hat{\omega}_n-\hat{\omega}_{n-1}P,\hat{\omega}_{n-1}\frac{\partial P}{\partial\theta_i}\right\rangle,\\
F_i(n,\tilde{\theta})&=\left\langle\hat{\omega}_n-\hat{\omega}_{n-1}\widetilde{P},\hat{\omega}_{n-1}\frac{\partial P}{\partial\theta_i}\bigg|_{\widetilde{P}}\right\rangle.
\end{align*}
The asymptotic normality of the least squares estimator $\widehat{P}_N$ is established in Theorem~\ref{asymnormal}.
\begin{thm}
\label{asymnormal}
Under Assumptions (A1)---(A4), 
\begin{align*}
N^{1/2}(\hat{\theta}_N-\tilde{\theta})\overset{\mathcal{L}}\to\mathcal{N}(0,A^{-1}\Sigma A^{-1}),
\end{align*}
where $\tilde{\theta}=\mbox{vec}(\widetilde{P})$, $\hat{\theta}_N=\mbox{vec}(\widehat{P}_N)$, and 
$$\Sigma_{ij}=E(F_i(0,\tilde{\theta}))F_j(0,\tilde{\theta}))+2\sum\limits_{k=1}^\infty E(F_i(0,\tilde{\theta})F_j(k,\tilde{\theta})).$$
\end{thm}
\begin{remark}The estimation of the transition probability matrix is consistent and asymptotically normal. Therefore, it is safe to use the SP method for prediction, as the estimation behaves stably with large sample size.
\end{remark}

\section{Simulations}
\label{s4}
Finite sample simulations were implemented to illustrate the effectiveness of the SP method. The method was tested on a FAR($1$) process with phase variation. In each simulation run, $300$ (or $600$) functions were simulated, and the first 90\% of the simulated trajectories were used for model fitting to do one-step ahead prediction for the remaining 10\% of trajectories by a moving-window approach. 
Each simulation run was repeated 200 times. 
We compared our method with the prediction method of Aue et al.~\cite{r3}, which does not incorporate functional registration, through two kinds of distance, the $\ell^2$ Euclidean distance ($\ell^2$) and the amplitude distance (FR). 
{We also compared our proposed state-space model with the transformation methods on the prediction of warping functions.}
\subsection{First simulation setup}
\subsubsection{Simulation of warping function}
\label{s411}
Based on the properties of $B$-splines (de Boor \cite{r6}), we develop the following procedure to simulate warping functions. We first generated four prototype warping functions with 7 $B$-splines. The $B$-spline scores of the four prototypes were generated through the following procedure:
{
\begin{enumerate}
\label{simulwarp}
\item  Four 6-variate vectors with positive elements, $\bm{\xi}_i=(\xi_{i1},\ldots,\xi_{i6}),\ i=1,2,3,4$, were specified as follows:
\begin{align*}
\bm{\xi}_1=(1.0, 1.2, 1.4, 1.6, 1.8, 2.0),\qquad\bm{\xi}_2=(2.0, 1.8, 1.6, 1.4, 1.2, 1.0),\\
\bm{\xi}_3=(0.3,0.3,1.2,1.2,0.3,0.3),\qquad\bm{\xi}_4=(1.2,1.2,0.3,0.3,1.2,1.2).
\end{align*}
\item  The vectors obtained in the first step were transformed as follows: $$\phi_{i,j+1}=\frac{\sum_{k=1}^j\xi_{ik}}{\sum_{k=1}^6\xi_{ik}},\qquad j=1,2,\ldots,6,$$ 
then concatenate a zero to each of the vectors $(\phi_{i2},\ldots,\phi_{i7})$ for $i=1,2,3,4$ to finalize the score vectors of prototype warping functions.
\end{enumerate}
Following the above procedure}, the four score vectors $\bm{\phi}_i=(\phi_{i1},\ldots,\phi_{i7})$ were constrained to satisfy $\phi_{i1}=0$, $\phi_{i7}=1$ and $\phi_{i1}<\phi_{i2}<\ldots<\phi_{i7}$. 
The prototypes were generated with 7 $B$-splines 
$b_i(t)=\sum_{j=1}^7\phi_{ij}B_j(t),~t\in[0,1].$
{The independent error warping functions, denoted by $\gamma_n^e$, were simulated through the same procedure, say, first simulate a 6-dimensional vector $\bm{\xi}_n^e$, then transform it to $\bm{\phi}^e_n$ and take the $B$-spline expansion.  The elements in $\bm{\xi}_n^e$ independently follow the uniform distribution $U[1,2]$. }The state of warping functions were simulated under a Markov process with four states, and the probability transition matrix has the representation
\begin{equation*}
P=\left(\begin{array}{cccc}
p&(1-p)/3&(1-p)/3&(1-p)/3\\
(1-p)/3&p&(1-p)/3&(1-p)/3\\
(1-p)/3&(1-p)/3&p&(1-p)/3\\
(1-p)/3&(1-p)/3&(1-p)/3&p\\
\end{array}\right).
\end{equation*}
Each state is associated with a prototype. The final warping functions were obtained by
$\gamma_n(t)=(1-\tau)b_{c^{(f)}_n}(t)+\tau \gamma^e_n(t),$
where $0<\tau<1$ determining the proportion of signal, and $c^{(f)}_n$ is the simulated state of the $n$-th warping function. Figure~\ref{warping1} displays the simulated warping functions and the prototypes.
\begin{figure}[ht]
\centering
\includegraphics[scale=0.41]{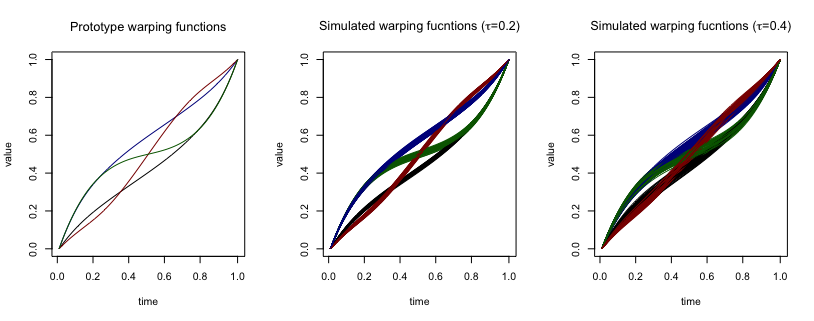}
\caption{Prototypes and simulated warping functions of different $\tau$'s}
\label{warping1}
\end{figure}
\subsubsection{Simulation of amplitude function}
Amplitude functions were simulated with the same 7 $B$-splines, where the scores of the third and the fifth B-splines are significantly larger than those of the other $B$-splines. Thus all curves have the same two-peak pattern. The two pronounced scores jointly follow a VAR($1$) process with varying coefficient matrix, and the amplitude functions were obtained by the basis expansion
$a_n(t)=\sum\limits_{j=1}^7\zeta_{nj}B_j(t).$
The VAR($1$) process has $4$ coefficient matrices specified below
\begin{align*}\left(
\begin{array}{cc}
0.8&0\\
0&0\\
\end{array}\right),
\left(
\begin{array}{cc}
0&0.8\\
0&0\\
\end{array}\right),
\left(
\begin{array}{cc}
0&0\\
0.8&0\\
\end{array}\right),
\left(
\begin{array}{cc}
0&0\\
0&0.8\\
\end{array}\right),
\end{align*}
and the coefficient matrices were determined by the state of warping function. The varying coefficient VAR(1) model is specified below,
\begin{align*}\left(
\begin{array}{c}
\zeta_{n+1,3}-4\\
\zeta_{n+1,5}-6
\end{array}\right)=\Phi^{(c^{(f)}_n)}\left(
\begin{array}{c}
\zeta_{n,3}-4\\
\zeta_{n,5}-6
\end{array}\right)+e_{n+1},
\end{align*}
where $e_n\overset{i.i.d}\sim\mathcal{N}(0,\Sigma)$, $\Sigma=0.2\bm{I}_2$, and $\bm{I}_2$ is a 2$\times$2 identity matrix. The other scores independently follow $\mathcal{N}(1,0.1)$.  The functional time series trajectories were obtained by applying the warping functions to the amplitude functions, $f_n(t)=a_n\circ\gamma_n(t).$

Figure~\ref{sim1} displays the simulated amplitude functions and the simulated functional time series for different $\tau$'s. Table~\ref{sim1t1} and \ref{sim1t2} display the average $\ell^2$ prediction error ($\ell^2$, defined as $\sum_{n}\|f_n-\widehat{f}_n\|_2/(0.1N)$) and amplitude difference (FR) between the predicted functions and the corresponding actual functions being predicted for $p=0.5,0.7,0.9$ and $N=300,600$.

\begin{figure}[!h]
\centering
\includegraphics[scale=0.4]{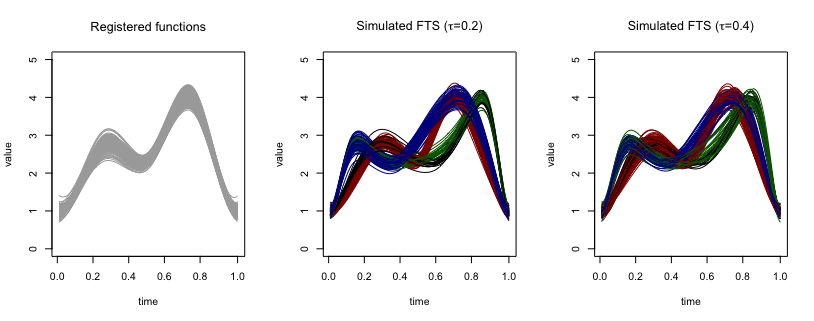}
\caption{Simulated curves for different $\tau$'s}
\label{sim1}
\end{figure}


\begin{table}
\centering
\captionof{table}{Average amplitude distance and $\ell^2$ prediction error ($\tau=0.4$). The format of each block: average (standard deviation$\times 100$)}
\begin{tabular}{|p{0.07in}|cc|cc|cc|}
\hline
\hline
   \multicolumn{1}{|c|}{$p$}& \multicolumn{2}{c|}{$0.5$}& \multicolumn{2}{c|}{$0.7$}& \multicolumn{2}{c|}{$0.9$}\\
\hline
 \multicolumn{7}{|c|}{Shape-preserving prediction}\\
\hline
 \multicolumn{7}{|c|}{$N$=300}\\
 \hline
$g$  & $\ell^2$ & \text{\rm FR} &$\ell^2$ & \text{\rm FR} &$\ell^2$ & \text{\rm FR}\\
\hline
	3&  0.298(1.86) &0.192(1.08) &0.257(2.37)& 0.199(1.12)&0.202(3.69)& 0.196(1.07)\\
	4&  0.294(1.75) &0.193(1.05) &0.241(2.46)& 0.199(1.02)&0.168(2.01)& 0.196(0.94)\\
	5& 0.297(1.79) &0.193(0.95)&0.245(2.36) &0.200(0.95)&0.170(2.29) &0.196(1.13) \\
\hline
 \multicolumn{7}{|c|}{$N$=600}\\
  \hline
$g$  & $\ell^2$ & \text{\rm FR} &$\ell^2$ & \text{\rm FR} &$\ell^2$ & \text{\rm FR}\\
\hline
	3& 0.300(1.26)& 0.189(0.75) &0.260(1.73)& 0.198(0.73)&0.200(2.50)& 0.194(0.85)\\
	4& 0.293(1.24)& 0.190(0.74)&0.242(1.69) &0.200(0.73)&0.166(1.51) &0.193(0.69)\\
	5& 0.295(1.31)& 0.191(0.71)&0.240(1.78) &0.199(0.71)&0.166(1.40) &0.193(0.78) \\
\hline
\hline
 \multicolumn{7}{|c|}{Prediction without registration}\\
 \hline
  \multicolumn{7}{|c|}{$N$=300}\\
 \hline
--  & $\ell^2$ & \text{\rm FR} &$\ell^2$ & \text{\rm FR} &$\ell^2$ & \text{\rm FR}\\
\hline
	--&  0.289(1.85)& 0.235(1.15) &0.245(2.12)&0.216(1.07)&0.185(2.23)&0.201(1.13)\\
	 \hline
  \multicolumn{7}{|c|}{$N$=600}\\
 \hline
--  & $\ell^2$ & \text{\rm FR} &$\ell^2$ & \text{\rm FR} &$\ell^2$ & \text{\rm FR}\\
\hline
	--&0.290(1.27)&0.236(0.83)&0.247(1.61)&0.214(0.80)&0.184(1.35)&0.199(0.80)\\
 \hline
\hline
\end{tabular}
\label{sim1t1}
\end{table}


\begin{table}
\centering
\captionof{table}{Average amplitude distance and $\ell^2$ prediction error ($\tau=0.2$). The format of each block: average (standard deviation$\times 100$)}
\begin{tabular}{|p{0.07in}|cc|cc|cc|}
\hline
\hline
   \multicolumn{1}{|c|}{$p$}& \multicolumn{2}{c|}{$0.5$}& \multicolumn{2}{c|}{$0.7$}& \multicolumn{2}{c|}{$0.9$}\\
\hline
 \multicolumn{7}{|c|}{Shape-preserving prediction}\\
\hline
 \multicolumn{7}{|c|}{$N$=300}\\
 \hline
$g$  & $\ell^2$ & \text{\rm FR} &$\ell^2$ & \text{\rm FR} &$\ell^2$ & \text{\rm FR}\\
\hline
	3&  0.377(2.18) &0.205(1.27) &0.318(3.03)& 0.202(1.22)&0.219(5.25)& 0.206(1.62)\\
	4& 0.371(2.25) &0.203(1.12)&0.291(3.14) &0.201(1.29)&0.166(2.40) &0.207(1.71)\\
	5& 0.371(2.25) &0.205(1.10)&0.292(3.41) &0.203(1.26)&0.168(2.46) &0.207(1.73) \\
\hline
 \multicolumn{7}{|c|}{$N$=600}\\
  \hline
$g$  & $\ell^2$ & \text{\rm FR} &$\ell^2$ & \text{\rm FR} &$\ell^2$ & \text{\rm FR}\\
\hline
	3& 0.381(1.39)& 0.206(0.81)&0.313(2.42) &0.202(0.99)&0.221(4.06) &0.209(1.24)\\
	4& 0.372(1.45)& 0.203(0.88)&0.289(2.14) &0.200(0.89)&0.169(1.92) &0.208(1.41)\\
	5& 0.373(1.65)& 0.203(0.90)&0.291(2.45) &0.201(0.86)&0.168(1.89) &0.207(1.40) \\
\hline
\hline
 \multicolumn{7}{|c|}{Prediction without registration}\\
 \hline
  \multicolumn{7}{|c|}{$N$=300}\\
 \hline
--  & $\ell^2$ & \text{\rm FR} &$\ell^2$ & \text{\rm FR} &$\ell^2$ & \text{\rm FR}\\
\hline
	--&  0.352(2.35)& 0.317(1.42)&0.285(2.71)&0.260(1.58)&0.179(2.77)&0.202(1.67)\\
	 \hline
  \multicolumn{7}{|c|}{$N$=600}\\
 \hline
--  & $\ell^2$ & \text{\rm FR} &$\ell^2$ & \text{\rm FR} &$\ell^2$ & \text{\rm FR}\\
\hline
	--& 0.354(1.45)&0.318(1.15)&0.279(2.31)&0.256(1.07)&0.176(1.95)&0.198(1.19)\\
 \hline
\hline
\end{tabular}
\label{sim1t2}
\end{table}


\subsection{Second simulation setup}
\label{s4.2}
In the second setup, the amplitude functions were simulated similarly with one coefficient matrix $\Phi=0.8\bm{I}_2$.  The major difference is the simulation of warping functions. In this simulation setup, the same procedure is applied to simulate a sequence of independent warping functions $\{\gamma_n^e(t)\colon n\in\mathbb{N}\}$, where ${\xi}^e_{n,j}\overset{i.i.d}\sim U[0.5,3]$ for $j=1,\ldots,6$, and then take the moving average of these functions to obtain $\{\gamma_n(t)\colon n\in\mathbb{N}\}$, say, $\gamma_{n}=\beta\gamma^e_{n-1}+(1-\beta)\gamma_{n}^e,$
where $\beta$ takes value in $(0.3,0.5, 0.7)$.
Here, the amplitude and warping functions were predicted separately.
Table~\ref{tab2} shows the average $\ell^2$ prediction error and amplitude distance between the predicted curves and the corresponding real curves for different values of $\beta$ and $N$.

\begin{table}[ht]
\centering
\captionof{table}{Average amplitude distance and $\ell^2$ prediction error for different $\beta$. The format of each block: average (standard deviation$\times 100$)}
\begin{tabular}{|p{0.07in}|cc|cc|cc|}
\hline
\hline
   \multicolumn{1}{|c|}{$\beta$}& \multicolumn{2}{c|}{$0.3$}& \multicolumn{2}{c|}{$0.5$}& \multicolumn{2}{c|}{$0.7$}\\
\hline
 \multicolumn{7}{|c|}{Shape-preserving prediction}\\
\hline
 \multicolumn{7}{|c|}{$N$=300}\\
 \hline
$g$  & $\ell^2$ & \text{\rm FR} &$\ell^2$ & \text{\rm FR} &$\ell^2$ & \text{\rm FR}\\
\hline
	4&  0.309(2.89)& 0.202(1.11) &0.280(2.55) &0.198(1.15)&0.310(2.88) &0.202(1.21)\\
	6&  0.309(2.67)& 0.203(1.10) &0.282(2.66) &0.199(1.09)&0.309(2.90) &0.204(1.11)\\
	8& 0.308(2.88) &0.205(1.05)&0.278(2.66) &0.201(1.16)&0.310(2.60) &0.204(1.07) \\
	10& 0.311(2.96) & 0.205(1.03)&0.279(2.60)& 0.201(1.11)&0.310(2.96)& 0.206(1.15) \\
\hline
 \multicolumn{7}{|c|}{$N$=600}\\
  \hline
$g$  & $\ell^2$ & \text{\rm FR} &$\ell^2$ & \text{\rm FR} &$\ell^2$ & \text{\rm FR}\\
\hline
	4& 0.310(1.99) &0.201(0.81) &0.279(1.92)& 0.197(0.72)&0.306(2.04) &0.200(0.82)\\
	6&  0.308(2.01)& 0.202(0.77) &0.276(1.88)& 0.197(0.71)&0.307(1.87)& 0.202(0.80)\\
	8& 0.304(2.03) &0.202(0.85)&0.279(1.85) &0.198(0.78)&0.306(1.95) &0.203(0.84) \\
	10&0.305(1.88)& 0.202(0.84)&0.277(1.91)& 0.197(0.69)&0.306(1.85)& 0.203(0.79) \\
\hline
\hline
 \multicolumn{7}{|c|}{Prediction without registration}\\
 \hline
  \multicolumn{7}{|c|}{$N$=300}\\
 \hline
--  & $\ell^2$ & \text{\rm FR} &$\ell^2$ & \text{\rm FR} &$\ell^2$ & \text{\rm FR}\\
\hline
	--&  0.301(2.76)& 0.262(1.60) &0.267(2.41)&0.234(1.49)&0.301(2.65)&0.261(1.64)\\
	 \hline
  \multicolumn{7}{|c|}{$N$=600}\\
 \hline
--  & $\ell^2$ & \text{\rm FR} &$\ell^2$ & \text{\rm FR} &$\ell^2$ & \text{\rm FR}\\
\hline
	--& 0.301(1.87)&0.261(1.12)&0.265(1.78)&0.232(0.98)&0.297(1.70)&0.260(1.09)\\
 \hline
\hline
\end{tabular}
\label{tab2}
\end{table}

\subsection{Discussion on the simulations}
In the first simulation setting, the optimal number of hidden states of warping functions is 4. Therefore, as $g$ changes from 3 to 4, the performance of the SP method is significantly improved. Table~\ref{sim1t1}---\ref{tab2} show that
(1.) The SP method preserves the common pattern better after incorporating functional registration into prediction, and
(2.) The performance of the SP method is robust to the selection of the number of hidden states. 
When the phase variation is difficult to  predict, the prediction by the SP method may not be as accurate as the prediction without functional registration. However, if the shape of the curve to be predicted is of major concern, the SP method is still a good alternative. 

\subsection{Comparison with logarithm transformation methods}
As has been discussed in the introduction, one feasible prediction approach for warping functions is predicting the transformed warping functions. Such methods typically transform highly constrained warping functions to unconstrained functions, and then linear models are employed to predict the transformed functions. The transformations in such methods always incorporate ``logarithm'' and the original variation is shrunk or exaggerated. To show the superiority of the state-space model approach, it was compared with two transformation methods.

The first competitor method employs Jupp transformation. The method was considered in the warped regression model (Gernivi \cite{r9}). In this method, each warping function $\gamma_n(t)$ is evaluated at fixed grids $0=\gamma_{n0}<\gamma_{n1}<\cdots<\gamma_{nr}<\gamma_{n,r+1}=1$, and each discretized vector is transformed by the Jupp transformation specified below:
$$\tau_{nj}=\log\left\{\frac{\gamma_{n,j+1}-\gamma_{n,j}}{\gamma_{n,j}-\gamma_{n,j-1}}\right\},\qquad j=1,\ldots,r,\ n=1,\ldots,N.$$
VAR model is then employed to predict the transformed vectors. The last step is to transform the predicted vector $(\hat{\tau}_{N+1,1},\ldots,\hat{\tau}_{N+1,r})$ back into a constrained increasing vector by the inverse Jupp transformation, say,
\begin{align*}
\hat{\gamma}_{N+1,j}&=s_{N+1,j}/(1+s_{N+1,r}),\\
s_{N+1,j}&=\sum_{k=1}^j\exp(\hat{\tau}_{N+1,1}+\cdots+\hat{\tau}_{N+1,k}),\qquad j=1,\ldots,r.
\end{align*}
This method typically requires fine grids so that the discretized vectors capture the major features of warping functions. 

The second method is a functional approach. Similar to those transformations employed in Petersen and M\"uller \cite{r23}, the transformation applied in this method has strict inverse only modulo the quotient space, and specifically in this method, two functions $f_1(t)$ and $f_2(t)$ defined over $[0,1]$ are equivalent if $f_1(t)/f_1(1)= f_2(t)/f_2(1) $. The transformation $\psi(\cdot)$ and its inverse are given as follows:
\begin{align*}
r_n(t)&\equiv\psi(\gamma_n)(t)=\log\left({\dot{\gamma_n}}\right)(t),\\
\psi^{-1}(r_n(t))&=\int_0^t\frac{\exp\left(r_n(x)\right)}{\int_0^1\exp(r_n(s))ds}dx.
\end{align*}
The prediction method proposed by Aue et al.~\cite{r3} was applied to predict the future transformed functions, and the prediction was then transformed back to a warping function with $\psi^{-1}(\cdot)$. 

The warping functions were simulated under the second setup ($\beta=0.8$), and $\{\gamma_i^e(t)\colon i\in\mathbb{N}\}$ were simulated in the same way (see Section~\ref{s411}) with 10 B-splines, say, $\gamma_i^e(t)=\sum\limits_{j=1}^{10}\phi_{ij}B_j(t).$ The scores $(\xi_{ij}\colon j=1,\ldots,9)$ follow the following distribution
\begin{align*}
&P(\xi_{ij}=1)=P(\xi_{ij}=2)=P(\xi_{ij}=3)=P(\xi_{ij}=4)=1/4,~j=1,2,3,\\
&P(\xi_{ij}=0)=q,P(\xi_{ij}=1)=(1-q)/2,P(\xi_{ij}=2)=(1-q)/2,~j=4,5,6,\\
&P(\xi_{ij}=1)=P(\xi_{ij}=2)=P(\xi_{ij}=3)=P(\xi_{ij}=4)=1/4,~j=7,8,9.
\end{align*}
Here, 500 warping functions were simulated. In the Jupp transformation method (JP), the warping functions were evaluated at 10 equally-spaced grids between 0 and 1. In the functional transformation approach (FT), 10 functional principal components were employed to represent the functions $\{r_n(t)\colon n=1,\ldots,500\}$. In our state space model method (MC), 10 prototypes were selected. The prediction error of $\hat{\gamma}(t)$ was evaluated with the spherical geodesic distance $d(\hat{\gamma}(t),\gamma(t))=\cos^{-1}(\langle S(\hat{\gamma}(t)),S({\gamma}(t))\rangle)$. Table~\ref{disad} displays the average prediction errors of different methods under different $q$'s.

\begin{table}[H]
    \centering
    \caption{Average prediction errors}
\begin{tabular}{|c|c|c|c|c|}
\hline
\hline
\diagbox [width=8em,trim=l] {Method}{$q$} & 0 & 0.3 & 0.5 & 1  \\
\hline
MC & $\bm{0.055}$ &$\bm{0.095}$  & $\bm{0.117}$& $\bm{0.045}$ \\
JP & 0.078 & 0.122 & 0.155&0.289\\
FT & 0.172 & 0.240&0.260 &0.326\\
\hline
\hline
\end{tabular}\vspace{0cm}
\label{disad}
\end{table}

As $q$ is large (close to 1), the middle part of the simulated warping functions is more likely to be flat, say, $\dot{\gamma}_n\approx0$, and the ``logarithm'' transformations are more likely to exaggerate the original variation. Table~\ref{disad} shows that the MC method is superior to the other two competitor methods, especially when the warping functions are flat over some intervals.

\section{Analysis of Pollution Concentration Trajectories}
\label{s5}
The SP method was applied to predict the air quality trajectories (De Vito et al.~\cite{r24}). The data is available at the \href{https://archive.ics.uci.edu/ml/datasets/Air+Quality#}{UCI Machine Learning Repository}. The dataset contains hourly averaged observations collected from 5 metal oxide chemical sensors embedded in an Air Quality chemical multi-sensor device. The device was placed at road level in a significantly polluted area in an Italian city. The pollution concentration was recorded from March 2004 to February 2005 (one year). Here we analyzed the Non Metanic Hydrocarbons concentration data since it contains less missing values. 

The pollution concentration is highly influenced by the traffic flow, so the trajectories share a common two-peak pattern (one peak in the morning and one peak in the afternoon). However, humidity, wind speed, temperature and other environmental factors can also influence the concentration, thus the trajectories display phase variation. Figure~\ref{pollution} displays the smoothed NMHC concentration trajectories and the registered trajectories. The trajectories of the weekdays are marked in black; those of Saturdays 
 are marked in blue; and those of Sundays are marked in red. Figure~\ref{pollutionwarping} displays the (prototype) warping functions.
\begin{figure}[!h]
\centering
\includegraphics[scale=0.28]{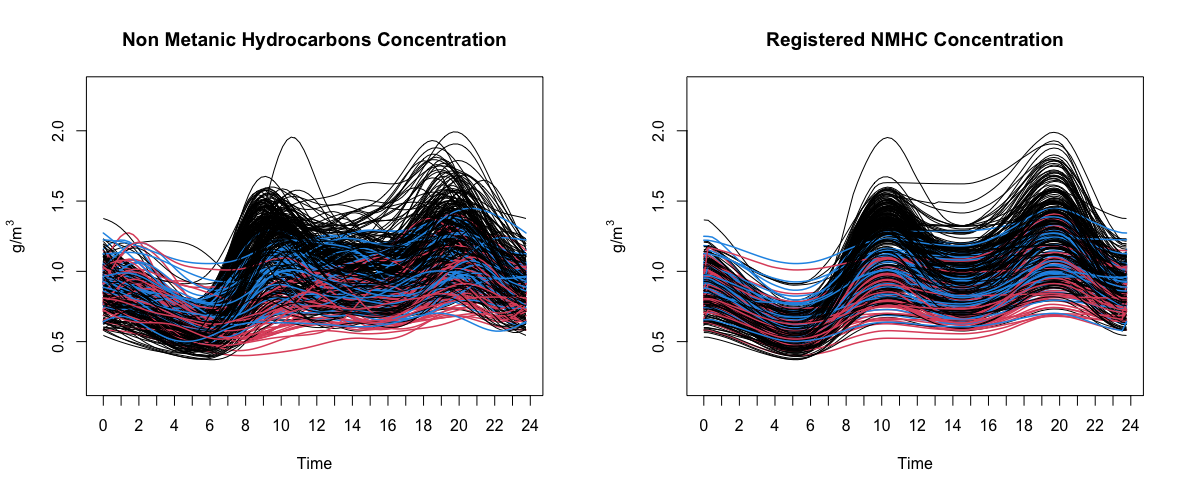}
\caption{Smoothed daily NMHC concentration and registered trajectories}
\label{pollution}
\end{figure}

As the trajectories for the weekdays share a different mean from that of weekends, the amplitude functions were centralized with the means of each day of the week. After removing the days with too many missing values, there are 357 trajectories in total. We found that the SP method produced the overall best prediction when the amplitude and warping functions are predicted separately. The first 300 curves were used to train the model for predicting the rest 57 curves, and FAR($1$) models were fitted to predict the amplitude functions of the trajectories to be predicted.  The SP method was compared with the prediction method without functional registration (Aue et al.~\cite{r3}). 
\begin{figure}[!h]
\centering
\includegraphics[scale=0.3]{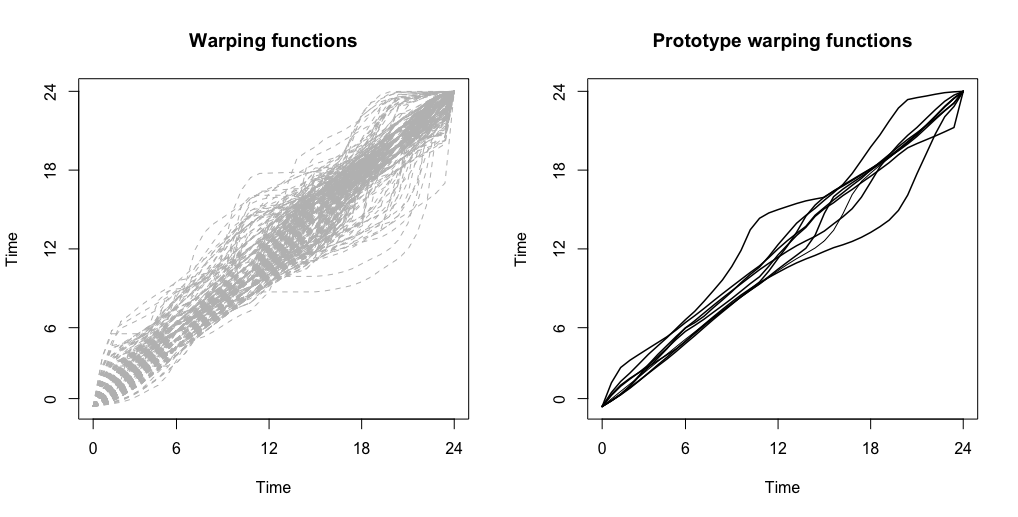}
\caption{Warping functions and prototype warping functions.}
\label{pollutionwarping}
\end{figure}

Table~\ref{comparison} displays the average $\ell^2$ prediction error and amplitude distance between the predicted curves and the corresponding smoothed curves under different numbers of states $g$ and dimensions of eigen-space $d$. It is noted that the SP method sacrifices marginal prediction accuracy to preserve the common two-peak pattern of the functional time series. The best SP prediction is achieved when $d=3,\ g=10$, while the competitor method reaches its best performance when $d=7$. This is because the functional registration step assures that less fPCs are needed to capture most of the vertical variation. 

\begin{table}
\centering
\caption{Prediction comparison. Format of each block: average $\ell^2$ prediction error (average amplitude distance)} 
\begin{tabular}{|c|ccc|}
\hline
\hline
 \multicolumn{4}{|c|}{Shape-preserving method} \\
 \hline
\diagbox [width=8em,trim=l] {$g$}{$d$}  & 3 &5 &7 \\
\hline
6&141.805(0.856)&145.567(0.848)&147.846(0.856)\\
7&141.988(0.854)&145.507(0.846)&148.507(0.866)\\
8&142.608(0.851)&144.627(0.857)&141.010(0.857)\\
9&140.398(0.846)&144.824(0.854)&143.171(0.848)\\
10&{\bf138.352(0.850)}&144.326(0.847)&145.268(0.847)\\
11&141.591(0.844)&146.282(0.847)&141.682(0.855)\\
\hline
\hline
\multicolumn{4}{|c|}{Prediction without functional registration} \\
\hline
$d$ & 3  &5 &7 \\
\hline
--&138.552(0.892)&136.458(0.879)&{\bf 134.998(0.880)}\\
\hline
\hline
\end{tabular}
\label{comparison}
\end{table}

\section{Conclusions}
\label{s6}
In this paper, we develop a new prediction method for stationary functional time series that display a common pattern. To the best of our knowledge,  our SP method is the first to incorporate functional registration into prediction of functional time series. The prediction algorithm jointly predicts the amplitude and phase components. These two predicted components are then combined to form the final prediction.

The SP method has two main advantages. First, if the curves displayed a common pattern and significant phase variation, considering vertical variation only would lead to the loss of main features. Comparatively, the new methodology separates amplitude and phase components first, thus the SP method can preserve the pattern better. Second, the SP method is ``natural'' in the following sense:
(1.) $S(\cdot)$ is a bijective transformation, thus no further adjustments are needed to transform a SRSF back to a warping function, which avoids bias; 
(2.)\  The method does not directly apply linear models to non-linear objects, making the prediction natural and avoiding extremely small values resulting from the ``logarithm''. The simulation study and real data analysis of Non Metanic Hydrocarbons Concentration data show that the SP method is superior to the prediction methods without registration in capturing the common pattern of trajectories, and meanwhile produce predictions with competitive prediction accuracy. In this paper, it is assumed that the pattern (shape) repeats with a fixed period. However, in some cases, such as some biomedical or physical signals, a signal may be composed of multiple components, and each component repeats itself at different rate.  The extension of the SP method to such cases is a research topic that will be pursued in the future.

\end{document}